# Antidepressant-like Effects of Neuropeptide SF (NPSF)

## Masaru Tanaka, Gyula Telegdy

*Abstract*—Neuropeptide SF (NPSF) is a member of RFamide neuropeptides that play diverse roles in central nervous system. Little is know about the effects of NPSF on brain functions. Antidepressant-like effect of NPSF was studied in modified mice FST. NPSF showed the antidepressant-like effects by decreasing the immobility time and increasing the climbing and swimming time. Furthermore, the involvement of the adrenergic, serotonergic, cholinergic or dopaminergic receptors in the antidepressant-like effect of NPSF was studied in modified mice FST. Mice were pretreated with a non-selective α-adrenergic receptor antagonist phenoxybenzamine, a β-adrenergic receptor antagonist, propranolol, a non-selective 5-HT$_2$ serotonergic receptor antagonist, cyproheptadine, nonselective muscarinic acetylcholine receptor antagonist, atropine, or D$_2$, D$_3$, D$_4$ dopamine receptor antagonist, haloperidol. The present results confirmed that the antidepressant-like effect of NPSF is mediated, at least in part, by an interaction of the α-adrenergic, 5-HT$_2$ serotonergic, muscarinic acetylcholine receptors and D$_2$, D$_3$, D$_4$ dopamine receptor in a modified mouse FST.

## I. INTRODUCTION

Neuropeptide SF (NPSF) is an 11-membered aminoacid peptide classified in RFamide neuropeptide family that shares RFamide motifs in the C-terminal. Five groups of RFamide neuropeptides include gonadotropin-inhibitory hormone (GnIH), neuropeptide FF (NPFF), pyroglutamylated RFamide peptide (QRFP), prolactin-releasing peptide (PrRP) and Kisspeptin [1]. RFamide neuropeptides are involved in a wide variety of physiological, neurophysiological and behavioral functions such as feeding, pain control and so on [2].

Antidepressant-like effects of neuropeptide AF (NPAF) and Kisspeptin were reported previously, but little is known about the functions of NPSF. NPSF was tested for antidepressant-like effects in modified forced swimming test (FST) in mice. Furthermore, the involvement of the adrenergic, serotonergic, cholinergic or dopaminergic receptors in the antidepressant-like effect of NPSF was studied in modified mice FST. Mice were pretreated with a non-selective α-adrenergic receptor antagonist phenoxybenzamine, a β-adrenergic receptor antagonist, propranolol, a non-selective 5-HT$_2$ serotonergic receptor antagonist, cyproheptadine, nonselective muscarinic acetylcholine receptor antagonist, atropine, or D$_2$, D$_3$, D$_4$ dopamine receptor antagonist, haloperidol.

## II. MATERIALS AND METHODS

### A. Animals

CFLP male mice were kept and handled during the experiments in accordance with the instructions of the University of Szeged Ethical Committee for the Protection of Animals in Research. Each animal was used in the experiments only once. The animals were housed in cages in a room maintained at constant temperature (25 ± 1 °C) and on a 12-h dark–light cycle (lights on at 06:00–18:00 h) with free access to tap water and standard laboratory food. Seven days of recovery from surgery was allowed before the experiments.

### B. Surgery

The mice were implanted with a cannula introduced into the right lateral brain ventricle in order to allow intracerebroventricular (*i.c.v.*) administration. The polystyrene cannula was inserted stereotaxically into the ventricle at the coordinates 0.2 mm posterior, 0.2 mm lateral to the bregma, and 2.0 mm deep from the dural surface [3]. The cannula was secured with cyanoacrylate (Ferrobond) (Budapest, Hungary). The mice were allowed seven days to recover from surgery before any i.c.v. administration.

### C. Materials

NPSF was purchased from Bachem (Bubendorf, Switzerland). Phenoxybenzamine hydrochloride was from Smith Kline & French (Herts, UK); propranolol hydrochloride was from ICI Ltd. (Macclesfield, UK); cyproheptadine hydrochloride was from Tocris (Bristol, UK); atropine was from EGYS (Budapest, Hungary); haloperidol was from G. Richter (Budapest, Hungary).

NPSF was lyophilized in a quantity of 10 μg per ampoule and stored at −20 °C Immediately before the experiments. NPSF was dissolved in sterile pyrogen-free 0.9% saline and administered i.c.v. via the cannula in a volume of 2 μl.

### D. Forced swimming test

The modified FST was conducted on the mice as reported previously [4]-[7]. The mice were forced to swim individually in a glass cylinder 12 cm in diameter and 30 cm

**Masaru Tanaka M.D., Ph.D.**, Institute of Pathophysiology, MTA-SZTE, Neuroscience Research Group, University of Szeged, Szeged, Hungary

**Gyula Telegdy, M.D., Ph.D., D.Sc.**, Institute of Pathophysiology, MTA-SZTE, Neuroscience Research Group, University of Szeged, Szeged, Hungary in height, filled with water to a height of 20 cm. The temperature of the water was adjusted to 25 ± 1 °C. The water was changed between the individual mice. A 15-min pretest session was followed 24 h later by a 3-min test session. NPSF (0.25 μg/2 μl, 0.50 μg/2 μl and 1.00 μg/2 μl i.c.v.) was administered 30 min before the test session. Physiological saline was used for the vehicle controls. A time-sampling technique was applied to score the durations of climbing, swimming and immobility. Climbing time was measured when the mouse was in an active vertical motion with its forelegs above the water level; swimming time was measured when the mouse was in horizontal motion on the surface of the water; and immobility time was measured when the mouse was in a upright position on the surface with its front paws together and making only those movements necessary to keep it afloat.

*E. Statistical analysis*

The two-way analysis of variance (ANOVA) test was followed by Tukey's test for multiple comparisons with unequal cell size. Probability values ($P$) of less than 0.05 versus the control and NPSF-administered group were considered significant

### III. RESULTS

NPSF (0.25 μg/2 μl, i.c.v.) administered mice exhibited a significantly decreased immobility time [$F(3, 49) = 17.31: P < 0.05$], a significantly increased climbing time [$F(3, 49) = 10.53: P < 0.05$] and a significantly increased swimming time [$F(3, 49) = 7.95: P < 0.05$]. NPSF (0.50 μg/2 μl, i.c.v.) administered mice exhibited a significantly decreased immobility time [$F(3, 49) = 17.31: P < 0.05$], a significantly increased climbing time [$F(3, 49) = 10.53: P < 0.05$] and a significantly increased swimming time [$F(3, 49) = 7.95: P < 0.05$]. NPSF (1.00 μg/2 μl, i.c.v.) administered mice exhibited a significantly decreased immobility time [$F(3, 49) = 17.31: P < 0.05$] and a significantly increased swimming time [[$F(3) = 7.95: P < 0.05$] (see Fig. I).

Phenoxybenzamine *per se* (2 mg/kg, *i.p.*) did not affect the immobility time, climbing time and swimming time. Pretreatment with phenoxybenzamine did not affect the immobility time the swimming time, but decreased the climbing time of NPSF-treated mice [$F(3.38) = 5.78: P < 0.05$] (see Fig. II).

Propranolol *per se* (5 mg/kg, *i.p.*) did not affect the immobility, climbing and swimming time and pretreatment with propranolol did not affect the immobility, climbing and swimming time of NPSF-treated mice time (data not shown).

Cyproheptadine *per se* (3 mg/kg, *i.p.*) did not affect the immobility, climbing and swimming time and pretreatment with cyproheptadine did not affect the immobility time of NPSF-treated mice. Pretreatment with cyproheptadine decreased the climbing [$F(3.40) = 8.12: P < 0.05$] and swimming time of NPSF-treated mice [$F(3.40) = 8.12: P < 0.05$] (see Fig. III).

Atropine *per se* (2 mg/kg, *i.p.*) did not affect the immobility, climbing and swimming time and pretreatment with atropinedid not affect the immobility time of NPSF-treated mice time. Pretreatment with atropine decreased the climbing [$F(3.40) = 5.80: P < 0.05$] and swimming time of NPSF-treated mice time [$F(3.40) = 4.30: P < 0.05$] (see Fig. IV).

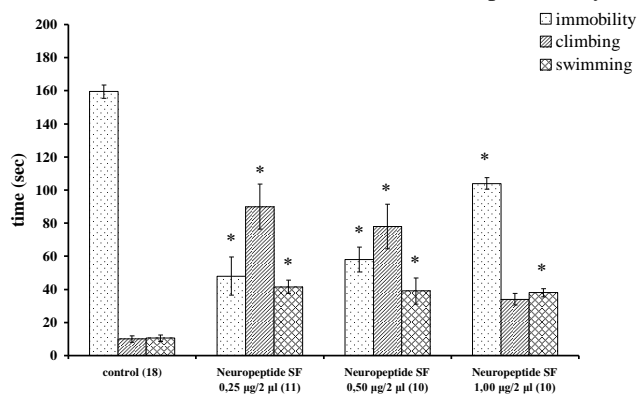

**Fig. I The antidepressant-like effects of Neuropeptide SF (NPSF) in modified mouse forced swim test (FST).** Control ($N$=18), NPSF 0.25 μg/2 μl, *i.c.v.* ($N$=11), NPSF 0.50 μg /2 μl, *i.c.v.* ($N$=10), NPSF 1.00 μg /2 μl, *i.c.v.* ($N$=10). * : $P < 0.05$ *vs.* control ($N$: the number of animals, $P$: probability)

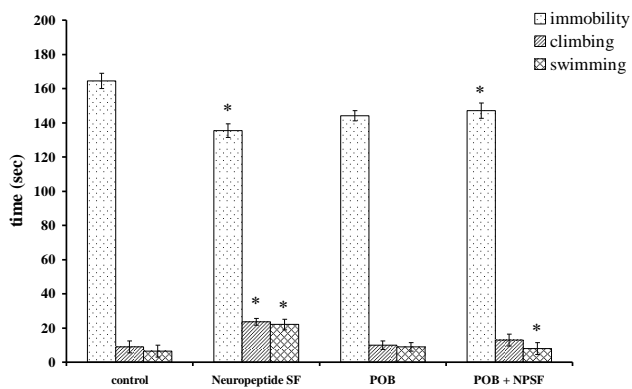

**Fig. II The effect of a nonselective α-adrenergic receptor antagonist, phenoxybenzamine (POB) in Neuropeptide SF (NPSF)-induced antidepressant-like action in modified mouse forced swim test (FST).** Control ($N$=9), Neuropeptide SF 0.25 μg/2 μl, *i.c.v.* ($N$=9), Phenoxybenzamine (POB) 2.0 mg/kg, *i.p.* ($N$=10), Phenoxybenzamine (POB) 2.0 mg/kg, *i.p.* + Neuropeptide SF 0.25 μg /2 μl, *i.c.v.* ($N$=10). *: $P < 0.05$ *vs.* control ($N$: the number of animals, $P$: probability).

Haloperidol *per se* (10 μg/kg, *i.p.*) did not affect the immobility, swimming and climbing time and pretreatment with haloperidol did not affect the climbing time of NPSF-treated mice time. Pretreatment with haloperidol increased the immobility time and decreased the swimming time of NPSF-treated mice time [$F(3.55) = 6.83: P < 0.05$] (Fig. 5).

The results demonstrated that the antidepressant-like effect of NPSF is mediated, at least in part, by an interaction of the α-adrenergic, 5-HT$_2$ serotonergic, muscarinic acetylcholine receptors and D$_2$, D$_3$, D$_4$ dopamine receptor in a modified mouse FST.

**Fig. III The effect of a nonselective 5-HT$_2$ serotonergic receptor antagonist, cyproheptadine (CPH) in Neuropeptide SF (NPSF)-induced antidepressant-like action in modified mouse forced swim test (FST).** Control (*N*=10), Neuropeptide SF 0.25 µg/2 µl, *i.c.v.* (*N*=10), Cyproheptadine (CPH) 3.0 mg/kg, *i.p.* (*N*=10), Cyproheptadine (CPH) 3.0 mg/kg, *i.p.* + Neuropeptide SF 0.25 µg /2 µl, *i.c.v.* (*N*=10). * : $P < 0.05$ vs. control (*N*: the number of animals, *P*: probability).

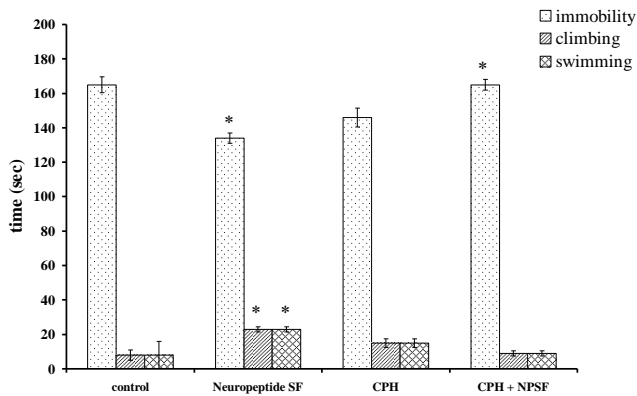

**Fig. IV The effect of a nonselective muscarinic acetylcholine receptor antagonist, atropin (ATR) in Neuropeptide SF (NPSF)-induced antidepressant-like action in modified mouse forced swim test (FST).** Control (*N*=10), Neuropeptide SF 0.25 µg/2 µl, *i.c.v.* (*N*=10), Atropin (ATR) 2.0 mg/kg, *i.p.* (*N*=10), Atropin (ATR) 2.0 mg/kg, *i.p.* + Neuropeptide SF 0.25 µg /2 µl, *i.c.v.* (*N*=10). * : $P < 0.05$ vs. control (*N*: the number of animals, *P*: probability).

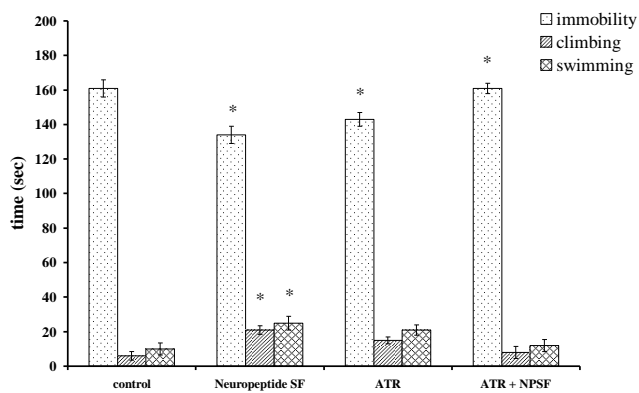

role in the control of energy metabolism and stress responses

**Fig. V The effect of a D$_2$, D$_3$, D$_4$ dopamine receptor antagonist, Haloperidol (HPD) in Neuropeptide SF (NPSF)-induced antidepressant-like action in modified mouse forced swim test (FST).** Control (*N*=13), Neuropeptide SF 0.25 µg/2 µl, *i.c.v.* (*N*=13), Haloperidol (HPD) 10 µg/kg, *i.p.* (*N*=14), Haloperidol (HPD) 10 µg/kg, *i.p.* + Neuropeptide SF 0.25 µg /2 µl, *i.c.v.* (*N*=15). * : $P < 0.05$ vs. control (*N*: the number of animals, *P*: probability).

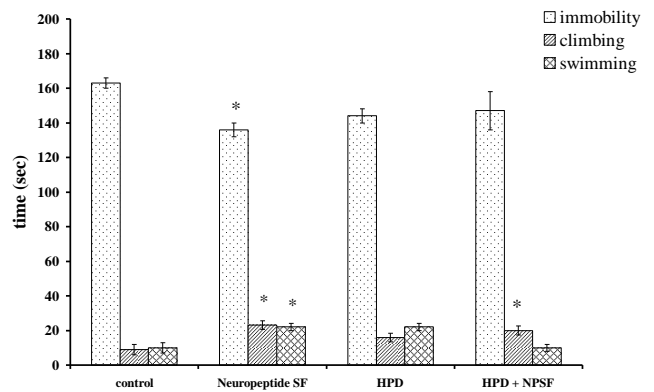

[10]. Furthermore, administration of RFRPs induces anxiety-related behavior in rats in open-field test. NPSF augments paraventricular CRH release and increases ACTH and corticosterone levels in the plasma [11].

Intracerebral administration of Kisspeptin-13 stimulates HPA axis and causes anxiety in rats [12]. Intracerebral administration of neuropeptide AF induces anxiety-like and antidepressant-like behavior [13]

The present study shows that NPSF has antidepressant-like effects in a modified FST of mice. Furthermore, the present results confirmed that the antidepressant-like effect of NPSF is mediated, at least in part, by an interaction of the α-adrenergic, 5-HT$_2$ serotonergic, muscarinic acetylcholine receptors and D$_2$, D$_3$, D$_4$ dopamine receptor in a modified mouse FST.

ACKNOWLEDGMENT

This work was supported by grants from TAMOP 4.2.1/B-09 and the Hungarian Academy of Sciences.

## IV. DISCUSSION

Since the first member of RFamide neuropeptides was discovered as cardioexcitatory peptide in invertebrates, many diverse group of N-terminal RFamide neuropeptides have been identified in the nervous system of higher animals [1]. Their roles in the central nervous system range from physiological processes to animal behaviors including feeding, pain control and so on [2].

Microinjection of RFRP-1 in the central nucleus of amygdala decreases food intake in the rat [8]. Central injection of neuropeptide AF reduces food intake in rats [9]. RFamides including prolactin-releasing peptide (PrRP) affect energy metabolism and neuroendocrine systems, playing a

REFERENCES

[1] G. K. Sandvik, K. Hodne, T. M. Haug, K. Okubo and F. A. Weltzien, "RFamide peptides in early vertebrate development," *Front Endocrinol. (Lausanne)*, vol. 5, 2014, p. 203.
[2] D. A. Bechtold and S. M. Luckman, "The role of RFamide peptides in feeding," *J. Endocrinol.*, vol. 192, 2007, pp. 3-15.
[3] A. S. Pellergrino and A. J. Cushman, *Stereotaxic Atlas of the Rat Brain*, New York, Plenum Press, 1979, pp. 8–57.
[4] M. J. Detke, M. J. Rickels and I. Lucki, "Active behaviors in the rat forced swimming test differentially produced by serotonergic and noradrenergic antidepressants," *Psychopharmacology (Berl.)*, vol. 121, 1995, pp. 66–72.
[5] R. T. Khisti, C. T. Chopde and S. P. Jain, "Antidepressant-like effect of the neurosteroid 3alpha-hydroxy-5alpha-pregnan-20-one in mice forced swim test," *Pharmacol. Biochem. Behav.*, vol. 67, 2000, pp. 137–143.
[6] T. Hokfelt, T. Bartfai and F. Bloom, "Neuropeptides: opportunities for drug discovery," *Lancet Neuro.*, vol. 2, 2003, pp. 463–472.

[7] R. D. Porsolt, M. Le Pichon and M. Jalfre, "Depression: a new animal model sensitive to antidepressant treatments," *Nature*, vol. 266, 1977, pp. 730–732.

[8] A. Kovács, K. László, R. Gálosi, K. Tóth, T. Ollmann, L. Péczely and L. Lénárd, "Microinjection of RFRP-1 in the central nucleus of amygdala decreases food intake in the rat," *Brain Res. Bull.*, vol. 88, 2012, pp. 589-595.

[9] B. A. Newmyer and M. A. Cline, "Neuropeptide AF is associated with short-term reduced food intake in rats," *Behav. Brain Res.*, vol. 219, 2011, pp. 351-353.

[10] Y. Takayanagi and T. Onaka, "Roles of prolactin-releasing peptide and RFamide related peptides in the control of stress and food intake," *J. Neuroendocrinol.*, vol. 23, 2011, pp. 20-27.

[11] M. Kaewwongse, Y. Takayanagi and T. Onaka, "Effects of RFamide-related peptide (RFRP)-1 and RFRP-3 on oxytocin release and anxiety-related behaviour in rats," *Regul. Pept.*, vol. 188, 2014, pp. 46-51.

[12] M. Jászberényi, Z. Bagosi, K. Csabafi, M. Palotai and G. Telegdy, "The actions of neuropeptide SF on the hypothalamic-pituitary-adrenal axis and behavior in rats," *Regul. Pept.*, vol. 188, 2014, pp. 46-51.

[13] M. Palotai, G. Telegdy, M. Tanaka, Z. Bagosi and M. Jászberényi, "Neuropeptide AF induces anxiety-like and antidepressant-like behavior in mice," *Behav. Brain Res.*, vol. 274, 2014, pp. 264-269.

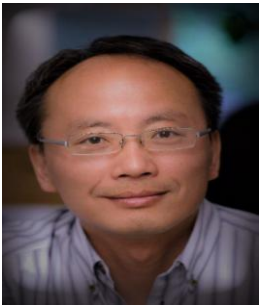

**Masaru Tanaka**, graduated from University of Szeged, Faculty of General Medicine, is currently working at Institute of Pathophysiology, Hungarian Academy of Science - University of Szeged (MTA-SZTE), Neuroscience Research Group, University of Szeged, Hungary. Main area of research is antidepressant effects of neurohormonal peptides. He published several research papers in Behavioral Brain Research, Brain Research Bulltin, Neuropeptides, Regular peptides, *etc*.